# Stabilization of hexazine rings in potassium polynitride at high pressure

Yu Wang[1], Maxim Bykov[2,3], Elena Bykova[2], Xiao Zhang[1], Shu-qing Jiang[1], Eran Greenberg[4], Stella Chariton[4], Vitali B. Prakapenka[4], Alexander F. Goncharov[1,2,*]

[1] Key Laboratory of Materials Physics, Institute of Solid State Physics, Chinese Academy of Sciences, Hefei 230031, Anhui, People's Republic of China

[2] Earth and Planets Laboratory, Carnegie Institution of Washington, 5251 Broad Branch Road NW, Washington, DC 20015, USA

[3] Department of Mathematics, Howard University, Washington, DC 20059, USA

[4] Center for Advanced Radiations Sources, University of Chicago, Chicago, Illinois 60637, USA

*Correspondence should be addressed to: agoncharov@carnegiescience.edu

Polynitrogen molecules represent the ultimate high energy-density materials as they have a huge potential chemical energy originating from their high enthalpy. However, synthesis and storage of such compounds remain a big challenge because of difficulties to find energy efficient synthetic routes and stabilization mechanisms. Compounds of metals with nitrogen represent promising candidates for realization of energetic polynitrogen compounds, which are also environmentally benign. Here we report the synthesis of polynitrogen planar $N_6$ hexazine rings, stabilized in $K_2N_6$ compound, which was formed from K azide upon laser heating in a diamond anvil cell at high pressures in excess of 45 GPa and remains metastable down to 20 GPa. Synchrotron X-ray diffraction and Raman spectroscopy are used to identify this material, also exhibiting metallic luster, being all consistent with theoretically predicted structural, vibrational and electronic properties. The documented here $N_6$ hexazine rings represent new highly energetic polynitrogens, which have a potential for future recovery and utilization.

Nitrogen-rich compounds have been widely sought as the great potential candidates for extremely high-yield high-energy density materials (HEDM) [1,2]. Their substantial potential energy storage or release capacity is related to a large energy difference of the single/double versus the triple bonds between nitrogen atoms. Numerous stable all-nitrogen compounds were predicted theoretically [3-7], however the existence of such materials at ambient conditions remain largely elusive— only a few radicals have been reported [8]. One major obstacle is the low-order N-N bonds tend to be unstable at low pressures, while they stabilize under pressure providing a pathway for synthesis. As a demonstration of a tour de force, the molecular diatomic nitrogen has been shown to transform into an atomic solid with single-bonded crystalline structure with a cubic gauche (cg-N) structure first via theoretical prediction at about 50 GPa [6], and then experimentally in the laser heated diamond anvil cell (DAC) at 110 GPa and 2500 K [9], but the recovery of this and other high-pressure polynitrogens remains problematic.

Various doping were thought for stabilization of N-rich compounds. Theoretical calculations predicted a number of high-pressure nitrogen-rich compounds with metals that form unusual low-bonded all-nitrogen groups as well as polymeric networks [10-16]. However, only few of them have been synthesized in the laboratory. Polynitrogens in the Mg-N and Fe-N systems containing $N_4^{2-}$ nitrogen chains or network were synthesized at about 50 and 100 GPa, respectively [17,18]; the cis-tetranitrogen $Mg_2N_4$ units are found to be recoverable at atmospheric pressure yielding great opportunities with regard to their possible use as HEDMs. Even more complex nitrogen rich compounds such as porous frameworks with transition metals (Re, Hf, W, and Os), where the structure combines nitrogen low-order bonded chains and triply bonded nitrogen molecules have been synthesized above 100 GPa [19,20]. However, stabilization of polynitrogen groups and large quantity production at ambient pressure remains a problem. For example, synthesis of stable *cyclo*- $N_5^-$ units at ambient pressure with a decomposition temperature below 117°C has required stabilization in very complex compounds such as, for example, $(N_5)_6(H_3O)_3(NH_4)_4Cl$ salt [21,22], at the expense of the energy density yield.

Alkali metals-nitrogen compounds were proposed to reduce the pressure of synthesis and to improve the compound stability and energy density of HEDMs as they are predicted to form a wealth of materials with various structures and compositions at high pressures [23-37]. The structural motifs include penta-$N_5^-$ salts, $N_6$ hexazine rings, and polymeric chains, and some of them are predicted to be preserved at ambient conditions. However, there are only a few experimental reports that support these findings. Steele *et al*. synthesized a penta-$N_5^-$ salts of Cs [30] by laser heating $CsN_3$ azide in molecular $N_2$ medium at 60 GPa in a DAC. The crystal structure of this $CsN_5$ with 6 formula units in the unit cell is very complex and it was determined by comparing the results of powder XRD and Raman spectroscopy with theoretical calculations. Laniel *et al*. synthesized $LiN_5$ compounds with penta-$N_5^-$ ions by heating Li in $N_2$ medium at 45 GPa, but the exact structural refinement could not be performed because of a powder nature of the sample [38,39]. They also documented the synthesis of Li diazenide $LiN_2$ at lower pressures. In a more recent work, Bykov *et al*. [40] synthesized two diazenide structures $NaN_2$ and $Na_3(N_2)_4$ containing dinitrogens $N_2^{-\delta}$ at 4 and 28 GPa, respectively, by laser heating $NaN_3$ azide. Use of azides containing linear $N_3^-$ groups as precursors allows to reduce substantially (*e.g.* compared to molecular $N_2$) the activation barrier for the reaction and provide more uniform reaction environment. Polymerization of various azides has been widely investigated in static DAC experiments[41,42], mainly at room temperature, and theoretically. The experiments show a rich polymorphism at high pressures, but the azide group remains (meta)stable [42-45] up to at least 60

GPa (in Li azide). However, Raman spectroscopy in NaN$_3$ azide above 19 GPa [41] shows an appearance of additional Raman bands at 1800-2000 cm$^{-1}$, which cannot be explained by the azide groups and likely indicate a chemical transformation as no associated phase transformation has been detected. It is interesting that the Raman bands in this spectral range can be also observed after a photochemical transformation at much lower pressures, 4.8-8.1 GPa, however, no XRD signature of the transformation could be found [46]. Here, we report a synthesis from KN$_3$ of previously predicted but never detected K$_2$N$_6$ compound which contains planar N$_6$ hexazine rings at about 50 GPa.

**Results**

The synchrotron X-ray diffraction (XRD) and Raman experiments in DAC are described in the **Methods** and **Supplementary Information**. Here, we focus on the experiments, where powdered KN$_3$ azide without any pressure medium/reagent was investigated, which gives a preference for formation of compounds with this composition. Additional experiments, where molecular N$_2$ was loaded together with KN$_3$ azide, are used to investigate the room temperature compression and low-temperature decompression behavior via *in situ* Raman measurements. When compressed at room temperature, KN$_3$ azide experiences phase transformations from *I*4/*mcm* to another structure at about 15 GPa [44,45], which has been predicted to be *C*2/*m* phase from theoretical calculations [29,32]. In this phase the linear N$_3$ azide groups are aligned parallel to each other unlike the low-pressure *I*4/*mcm*, where they are alternatively rotated by 90° with respect to each other. Our Raman data (Fig. S1 of Supplementary Information) do show a phase transition above 15 GPa manifested by splitting of the librational modes and appearance of new peaks in good agreement with Ref. [44]. Above 30 GPa, the Raman spectra show a broad band near 1900 cm$^{-1}$ increasing intensity with pressure similar to that reported previously in NaN$_3$ [41]. Our XRD data at ~50 GPa, which generally agree with the extrapolated results of Ref. [45], reveal a poorly crystallized single crystal of KN$_3$ with extremely broad diffracted peaks (see Supplementary Information, Fig. S2). The unconstrained lattice determination results in the following unit-cell parameters $a$ = 5.4252(13), $b$ = 5.487(11), $c$ = 4.67(4) Å, $\alpha$ = 90.6(4)°, $\beta$ = 92.2(4)°, $\gamma$ = 90.22(17)°. This indexing suggests that this phase is just a heavily distorted $I4/mcm$ ambient pressure structure. Indeed, the refinement against single-crystal data using the undistorted $I4/mcm$ model results in the $R_1$ factor of ~11%, which clearly indicates that no major structural rearrangements take place, thus ruling out the predicted $C2/m$ phase. The best refinement was obtained in a monoclinic space group $I2/c$ (Fig. S3, Table S1, Supplementary Information).

It is worth noting that KN$_3$ is sensitive to the X-ray exposure as evidenced by a darkening of the sample (Fig. 1c), a distinct Raman spectrum (Fig. 2), and time dependent XRD patterns (Fig. S4). The irradiated product demonstrates a substantial reduction N$_3$ azide modes and sharpening and strengthening of the Raman peaks, which could be assigned to yet unidentified definitely polynitrogen configuration. However, time-dependent X-ray diffraction measurements do not show the appearance of new diffracted peaks suggesting that the X-ray-induced phase is amorphous in nature (Fig. S4, Supplementary Information).

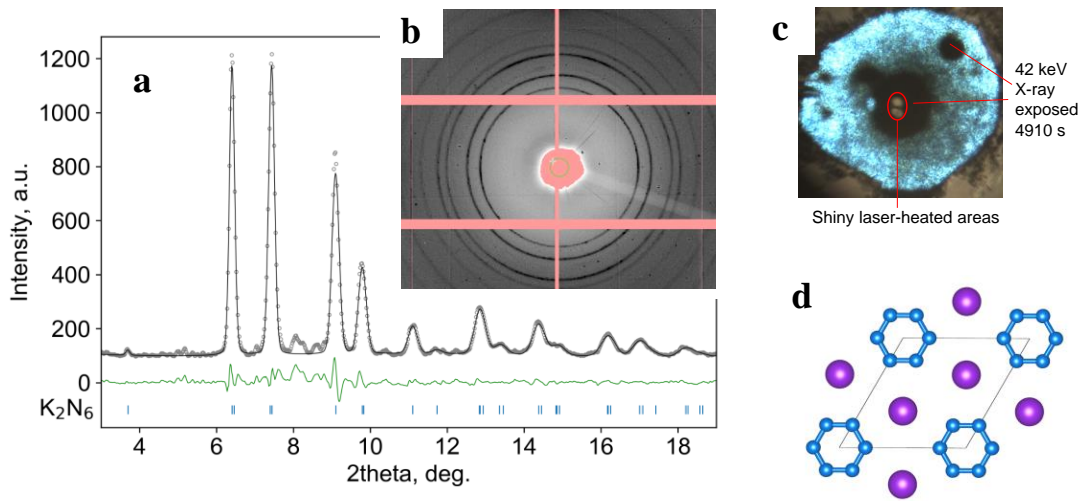

**Figure 1. X-ray diffraction data in compressed to 50 GPa and laser heated via direct coupling KN$_3$ without a pressure medium.** A freshly chosen (no X-ray exposure) spot was laser heated (**c**), which results in the appearance of the heated spot (circled by red) (inset). X-ray exposed areas that have been created around the heated spot while XRD mapping the area and in a control spot are labeled. The panel (**a**) shows a representative integrated 1D pattern in the laser heated location; the circles are the data and the line is the Le Bail refinement in the structure shown as a projection along the [001] direction in the panel (**d**). A few weak non-indexed peaks (*e.g.*, near 8°) correspond to the diffraction pattern of the non-reacted sample (Fig. S2). The panel (**b**) shows a 2D diffractogram. The X-ray wavelength is 0.2952 Å.

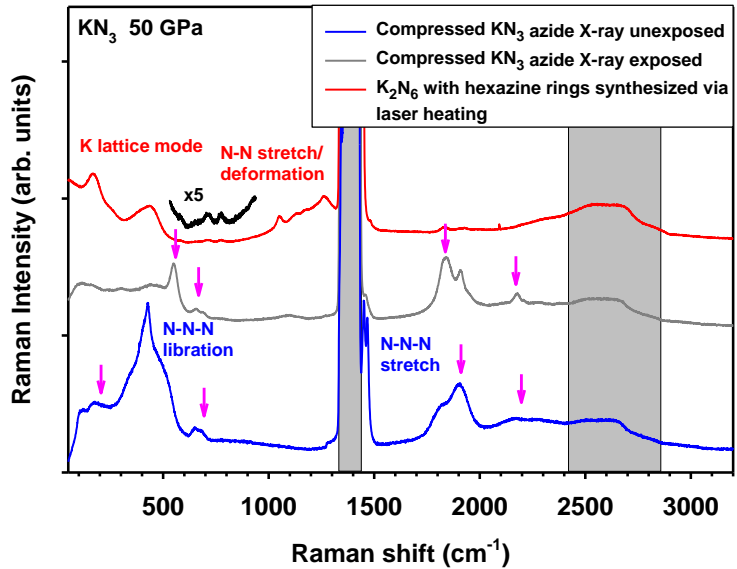

**Figure 2.** Raman spectra of $KN_3$ before (X-ray unexposed and exposed) and after laser heating at around 50 GPa. The pink arrows show the Raman peaks that emerge under pressure and due to X-ray radiation and are assigned to yet unidentified polynitrogen species. The grayed areas indicate the spectral ranges, where Raman peaks of the stressed diamond anvils appear.

Theoretical structure search calculations predict different than $I4/mcm$ $KN_3$ polymorphs to be stable above 15 GPa[27,29,32,37], however the reconstructive phase transitions are hindered due to the high kinetic barriers. In order to overcome these restrictions, we performed high-temperature treatment of cold-compressed $KN_3$. Laser heating of $KN_3$ at 49-53 GPa to 2500 K results in formation of a new material clearly identified by shiny surfaces and a very distinct XRD pattern (Fig. 1). The diffraction pattern indicates the synthesized material is a fine grain powder, which is well suited for powder diffraction methods. The unit cell was indexed with hexagonal unit cell ($a$ = 5.2841(9), $c$ = 2.6208(8) Å) (Fig. 1). This lattice is in a good agreement with the predicted high-pressure polymorphs of $KN_3$ – $K_2N_6$ with *P*6/*mmm* symmetry[29,32,37]. In this structure potassium and nitrogen atoms occupy Wyckoff sites 2*d* (2/3, 1/3, 1/2) and 6*j* (0, y, 0) respectively. Nitrogen atoms are connected with each other forming planar $N_6$ rings (Fig. 1). The theoretical unit cell parameters slightly differ (*a* is 5.5 % larger, while *c* is 4% smaller) from the experimentally determined ones yielding the volume discrepancy within fair 6% (Fig. 3).

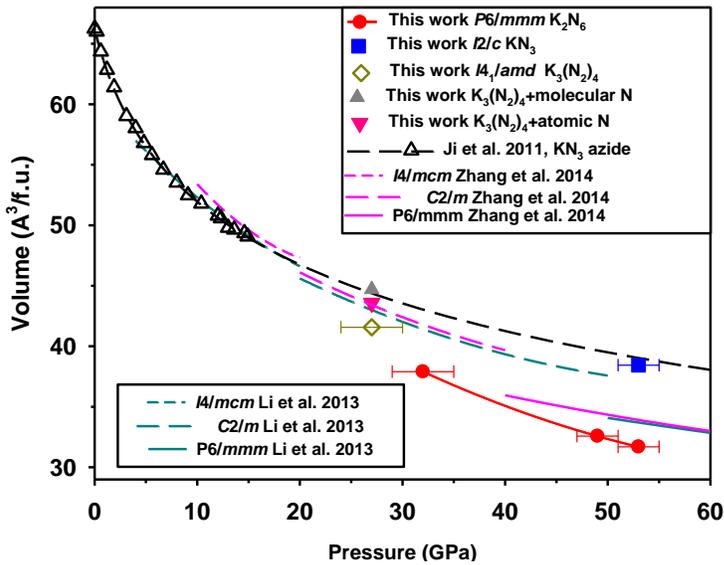

**Figure 3. Specific volumes of K-N compounds investigated here.** $KN_3$ azide, the *I*4/*mcm* structure of which was determined up to 15 GPa [45], develops a monoclinic distortion with the space group *I*2/*c*, determined at 53 GPa (Figs. S2, S3, Table S1), where upon laser heating it transforms to *P*6/*mmm* $K_2N_6$ compound with $N_6$ hexazine rings, which is some 17% denser. However, $I4_1/amd$ $K_3(N_2)_4$ (which is slightly deficient in nitrogen) synthesized at 30 GPa also assisted by laser heating has about the same density as *I*2/*c* azide. To take into account the difference in composition we added a specific volume of nitrogen in molecular and monatomic (cg-N) compounds at the same pressure determined from their equations of states [47,48]. The theoretical calculations of the specific

volumes of *I*2/*c*, *C*2/*m* (which we do not observe), and *P*6/*mmm* $K_2N_6$ are shown for comparison [29,32].

Considering the oxidation state of K as +1, each $N_6$ ring must accommodate 2 electrons leading to an 8 π-electron system in $N_6^{2-}$. Therefore, this system does not follow the 4*n*+2 Huckel rule for being an aromatic compound and 2 extra electrons are entering antibonding partially filled π* orbitals leading to metallic character of this compound (Fig. 1(c)). Theoretical calculations [29,32,37] predicted that *P*6/*mmm* $KN_3$ would be a stable polymorph above 40-50 GPa, which is consistent with our observations. However, there is a large kinetic barrier for such transformation as the system changes the chemical state. Thus, laser heating is needed for realization of this transformation, while mere compression to higher pressure would largely preserve the metastable azide phase as has been also demonstrated previously in Na [41] and Li [42] azides.

      The Raman spectra change drastically upon the transformation as seen in Fig. 2, which also shows that X-ray irradiated $KN_3$ also has a distinct Raman spectrum. The newly synthesized *P*6/*mmm* $KN_3$ shows a distinct Raman bands in the spectral range characteristic for the stretching vibrations of a low-bonded nitrogen (1000-1300 $cm^{-1}$), and two low-frequency modes in the spectral range of the N-N bending modes and lattice vibrations. Also weak peaks at about 700 $cm^{-1}$ can be seen. We have been able to trace the Raman spectrum of this phase down to approximately 20 GPa (Fig. S5) on unloading thus establishing their pressure behavior and allowing to compare the results with theoretical calculations of vibrational frequencies of *P*6/*mmm* $KN_3$ at 20 GPa [32] and 100 GPa [29] (Fig. 4) and a hypothetical $N_6$ hexaazabenzene isomer where the ring is out-of-plane distorted with the $D_2$ symmetry [7]. These data provide very important clues for assigning the observed Raman modes of newly synthesized *P*6/*mmm* $KN_3$ (Fig. 4).

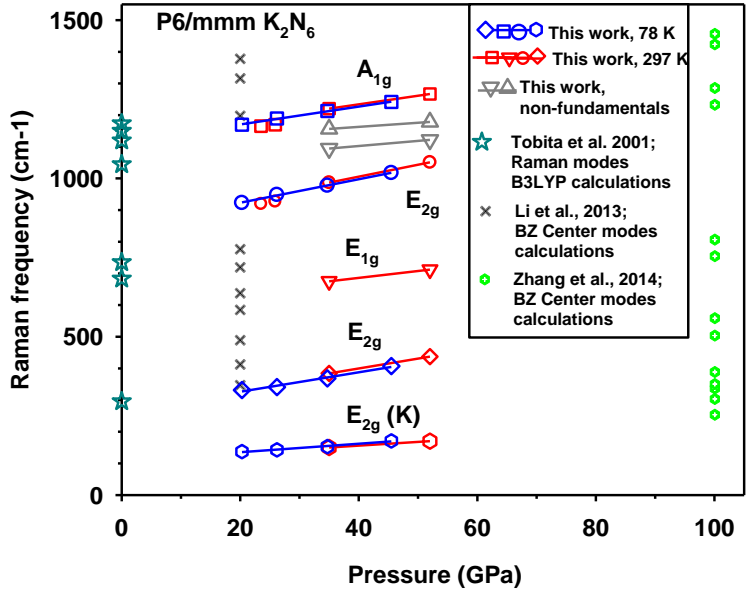

**Figure 4**. Vibrational frequencies of $P6/mmm$ $K_2N_6$ deduced from the Raman spectra of this work in comparison to the theoretically calculated Brillouin zone center modes of $P6/mmm$ $K_2N_6$ [29,32] and the Raman-active modes of a $N_6$ ring isomer in the twist-boat $D_2$ structure [7].

The group theory predicts four $A_{1g}$+ $2E_{2g}$ +$E_{1g}$ Raman active modes of the hexazine rings and one $E_{2g}$ mode for K vibrations; no translation or rotation modes the hexazine rings is expected because there is only one ring in the unit cell. The lowest-frequency mode corresponds to $E_{2g}$ vibrations of K atoms in the *xy* plane. This assignment is supported by the fact that a similar mode of the isostructural $Na_2N_6$ compound (this will be presented elsewhere [40]) has a substantially lower frequency, while the spectral positions of the other modes that belong to the hexazine ring are approximately the same. The second in frequency strong mode at approximately 400 cm$^{-1}$ is assigned to the $E_{2g}$ in-plane deformation mode while the next in frequency weak peak at 700 cm$^{-1}$ – to the $E_{1g}$ out-of-plane deformation mode of the hexazine ring. This assignment is based on the frequency sequence of the deformation bands calculated in Ref. [7] determined based on their symmetry correlated to the symmetry of the symmetric $D_{6h}$ planar ring; we also assumed that the $E_{2g}$ in-plane deformation mode must be stronger as it modulates intra-ring N-N bonds while the $E_{1g}$ out-of-plane mode is purely deformational. The two other in-plane vibrations at higher frequencies also yield strong Raman bands corresponding to the $E_{2g}$ in-plane mixed vibration-deformation mode and the symmetric breathing $A_{1g}$ stretching mode; the latter is expected to be the dominant Raman band (Fig. 2). There are weaker and broader Raman peaks between these modes at ~1150 cm$^{-1}$ (Fig. S6), which cannot be assigned to any Raman active fundamental vibrations assuming the $D_{6h}$ symmetry of the planar $N_6$ ring. However, the other vibration-deformation modes, which are nominally IR active, can contribute to Raman processes in the case of the symmetry reduction (*e.g.* dynamical), for example, to the nonplanar twist-boat $D_2$ structure [7]. The frequencies of the characteristic hexazine rings N-N stretching-deformation modes decrease with pressure release and these dependencies extrapolate reasonably well to the theoretically calculated values of the twist-boat $N_6$ isomer at ambient pressure [7].

On unloading of $K_2N_6$ with hexazine rings at 78 K the Raman spectra show clear signs of deterioration of this compound at 22 GPa, where the characteristic bands become weaker and other broad bands at 400, 1800, 2100, and 2300 cm$^{-1}$ increase in intensity (Fig. S5). The product is difficult to identify; it is definitely not azide as there is no sign of the characteristic N-N stretching mode of the linear $N_3$ groups. Based on observations of a number of high-frequency modes this product is likely a highly disordered state with elongated $N_2$ dimers. The Raman spectrum of this material is qualitatively similar to that of the cold compressed azide (Fig. 2).

Laser heating to 2500 K of unexposed $KN_3$ azide, first pressurized to 50 GPa and then unloaded at room temperature to 30 GPa, results in formation of a shiny material, which visually looks similar to *P6/mmm* $K_2N_6$ but is different in structure and composition judging from single-crystal XRD and Raman spectroscopy measurements (Fig. 5). The reaction products tend to be textured and single crystals of good quality and sufficient sizes can be selected by mapping the heated sample area. The structure and composition of the synthesized material was directly solved to be $I4_1/amd$ $K_3(N_2)_4$, which is isostructural to $Na_3(N_2)_4$ (Table S2), that has been synthesized in similar P-T conditions in our previous work [40].

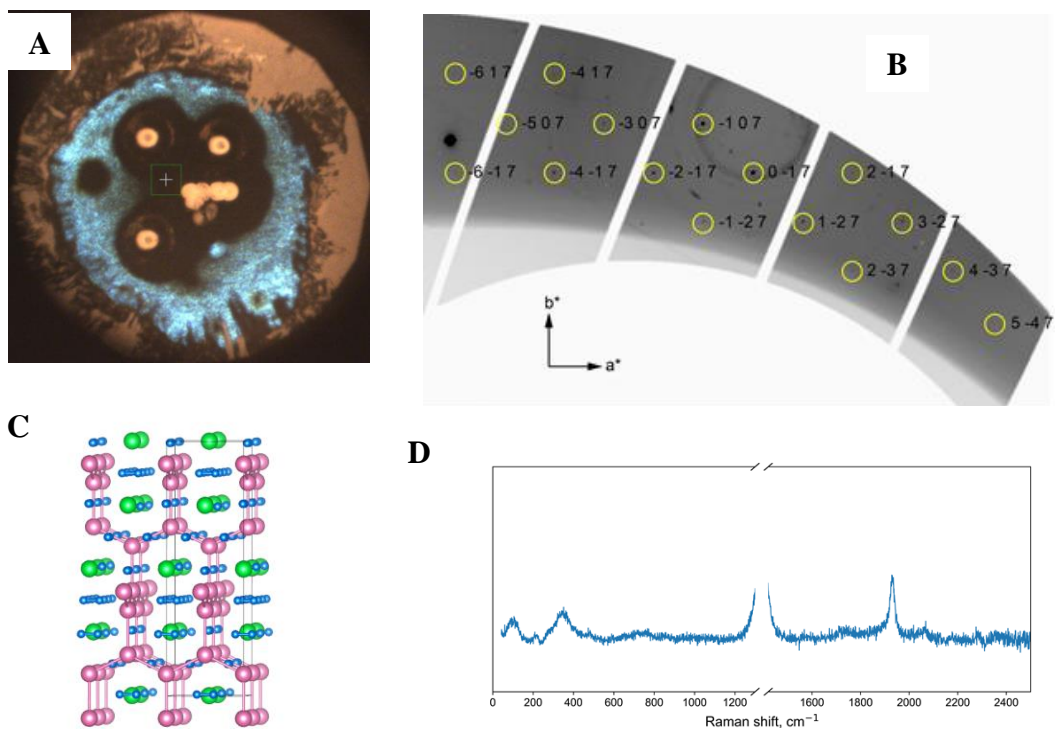

**Figure 5. Experimental data on $I4_1/amd$ $K_3(N_2)_4$ synthesized at 30 GPa by laser heating $KN_3$.** (A) A microphotograph of $KN_3$ sample laser heated in several places revealing shiny areas, where a new material is synthesized. Transparent parts of the sample (unexposed $KN_3$ azide) appear light blue in the transmitted light. (B) A part of the reconstructed ($hk7$) precession image from the single-crystal XRD dataset of $K_3N_8$ at ~27 GPa. Reflections $hkl$ with $h+k+l = 2n+1$ are absent due to the *I*-centering of the lattice. (C) The structure of *I*4$_1$/*amd* K$_3$(N$_2$)$_4$, which consists of two K sublattices (pink and green spheres) and two elongated N$_2$ dimers sublattices (blue spheres). (D) The Raman spectra of K$_3$N$_8$ at ~27 GPa. A high-frequency band corresponds to the N-N stretching mode of the N-N dimers. The low-frequency modes are vibrations of K atoms and lattice vibrations of N$_2$ dimers.

**Discussion**

Previous experiments on various azides at high pressures showed the stability of the linear N$_3^-$ ion to at least 60 GPa judging from the vibrational spectroscopy [41-45,49]. Theoretical calculations predict a major phase transformation to a monoclinic *C*2/*m* within this chemical structure, where the linear N$_3^-$ groups become collinear forming the *C*2/*m* structure[29,32]. Lithium azide, LiN$_3$, at ambient pressures and α-NaN$_3$ at 0.3-17 GPa has the same structure [31,43]. However, our experiments show that in K azide this structure cannot be easily realized. At room temperature, KN$_3$ is reported to transform to another phase with the linear N$_3^-$ groups [44,45] but the structure of this phase is not monoclinic *C*2/*m*. Instead our data (Figs. S2-S4) show that it rather remains in an only slightly distorted *I*4/*mcm* structure preserving the original lattice to at least 50 GPa. However, the azide ion becomes chemically unstable in this regime as manifested by formation of other yet unidentified polynitrogen species above 30 GPa (evidenced via emergence of a 1900 cm$^{-1}$ Raman band, Fig. S1), the process, which is accelerated by X-ray radiation exposure.

These experiments show that the azide ion N$_3^-$ configuration becomes unstable at high pressures. Our high-pressure laser heating experiments revealed two new classes of compounds that become stable at these conditions. The structural and vibrational spectroscopy data provide concerted consistent data identifying their physical and chemical structure.

Laser heating of $KN_3$ azide at 30 GPa results in formation of an unexpected diazenide $I4_1/amd$ $K_3(N_2)_4$ with a complex structure, which contains K atoms and elongated $N_2$ dimers with different site symmetries, and a very large unit cell. It is interesting that this structure does not seem to reveal more efficient packing than azide as the specific volume of $K_3(N_2)_4$ is very similar to that of $KN_3$ azide (Fig. 3). It is likely that realization of this structure is at least partially due to its high entropy stemming from rotating dimers. In contrast, the formation of $P6/mmm$ $K_2N_6$ with hexazine rings at 50 GPa is accompanied by a large volume contraction (about 17%), which explains the importance of pressure in its synthesis. This transformation manifests realization of the long sought low-bonded polynitrogen species at relatively low pressure compared to monatomic polymeric nitrogen.

To create compounds with other than explored here polynitrogen species, reagent compositions must be varied in the synthesis, for example by providing an excess of nitrogen. Our preliminary experiments show that laser heating experiments in $KN_3$ in $N_2$ medium results in a wealth of new compounds with different composition and structure, which contain various polynitrogen species; these results will be presented elsewhere.

In conclusions, our experiments show the azide group $N_3$ is prompt for chemical destabilization under pressure evidenced by a loss of crystallinity of the compressed at 300 K sample, spontaneous emergence of extra Raman bands, and sensitivity to X-ray radiation. The reactions product is likely comprises chemically modified nitrogen species such as, for example, $N_2$ dimers and/or square plane $D_{2h}$ $N_4^+$ [41], however, lack of crystallinity does not allow to identify them. Depending on pressure of the synthesis, two new crystalline materials form upon the laser heating: hexazine ring bearing $K_2N_6$ above 45 GPa and unusual $K_3(N_2)_4$ diazenide at 30 GPa. These compounds reveal a large metastability domain on unloading, however they are not stable at ambient conditions. The high-pressure synthesis of new polynitrides from azides has a lot of potential due to a reduced pressure of synthesis and a wealth of new compounds with a variety of chemical species and compositions that we documented here and others that need to be explored.

**Methods**

The high-pressure high-temperature behavior of $KN_3$ was studied on one sample intended for XRD measurements and another four samples for Raman measurements. A powder sample of potassium azide $KN_3$ was placed in sample chambers of DAC (see Supplementary information) equipped with diamond anvils with the culets of 250-300 μm diameter for single-crystal and powder diffraction XRD and Raman experiments. Re foil preindented to a thickness of 30 μm served as a gasket. Ruby chips were placed inside the sample chambers for pressure measurement. The XRD sample had no pressure-transmitting medium. All of the Raman samples were loaded with the nitrogen gas at high pressure of 0.15 GPa at room temperature. Raman samples #1 and #2 were quenched to below 2 GPa at room temperature while samples #3 and #4 were quenched to below 3 GPa at liquid nitrogen temperature using a continuous flow He/$N_2$ cryostat. All samples were compressed up to the target pressures and laser-heated ($\lambda = 1064$ nm) using double-sided laser-heating systems of the beamline GSECARS (APS, Argonne, USA) or ISSP (Hefei, China) (Supplementary Information). Laser heating of $KN_3$ in the DAC was performed without any absorber by directly coupling near IR laser radiation with the sample. $KN_3$ is a weak absorber at 50-60 GPa, however it can be coupled to the laser via grain boundaries and other defects, although care must be taken as it tends to run away once the temperature becomes higher than approximately 1000 K. Thus, the temperature at which the reported below reactivity is reported should be considered cautiously. The heated area becomes nontransparent; we performed XRD and Raman mapping of these areas where available.

XRD measurements were performed at the beamline GSECARS (13IDD, APS, Argonne, USA). We used monochromatic X-ray beam ($\lambda$ = 0.2952 Å) focused down to 3×3 µm$^2$ by a Kirkpatrick-Baez mirror system and diffraction patterns were collected on a Pilatus 1M detector (CdTe). For the single-crystal XRD measurements samples were rotated around a vertical $\omega$-axis in a range ±35°. The diffraction images were collected with an angular step $\Delta\omega$ = 0.5° and an exposure time of 1s or 2s/frame. For the details about the data analysis and structural solution, please see Supplementary Information.

Raman spectra of the sample studied at GSECARS using XRD combined with laser heating were measured concomitantly using the GSECARS Raman system with the excitation wavelength of 532 nm in the spectral range of 10 to 4000 cm$^{-1}$ with a 4 cm$^{-1}$ spectral resolution. The Raman spectra of the laser heated at ISSP samples were examined with 532 and 660 nm excitation lines using a similar custom system coupled to a continuous flow cryostat (See Supplementary Information for more details).

**Acknowledgements**

This work at ISSP was supported by the National Natural Science Foundation of China (Grant Nos. 11504382, 21473211, 11674330, 51672279, 11874361, 11774354, and 51727806), the CASHIPS Director's Fund (Grant No. YZJJ201705), the Chinese Academy of Science (Grant Nos. YZ201524 and YZJJ2020QN22), and a Science Challenge Project No. TZ201601. A.F.G. was partially supported by the Chinese Academy of Sciences Visiting Professorship for Senior International Scientists Grant No. 2011T2J20 and Recruitment Program of Foreign Experts.

This research at Carnegie and APS was sponsored by the Army Research Office and was accomplished under the Cooperative Agreement Number W911NF-19-2-0172. Concomitant Raman spectroscopy experiments were performed at GeoSoilEnviroCARS (The University of Chicago, Sector 13), Advanced Photon Source (APS), Argonne National Laboratory. GeoSoilEnviroCARS is supported by the National Science Foundation - Earth Sciences (EAR - 1634415) and Department of Energy GeoSciences (DE-FG02-94ER14466). The Advanced Photon Source is a U.S. Department of Energy (DOE) Office of Science User Facility operated for the DOE Office of Science by Argonne National Laboratory under Contract No. DE-AC02-06CH11357.

**Data availability statement**

The datasets generated during and/or analyzed during the current study are available from the corresponding author on reasonable request.

# Supplementary information to

# Stabilization of hexazine rings in potassium polynitride at high pressure

Yu Wang[1], Maxim Bykov[2], Elena Bykova[2], Xiao Zhang[1], Shu-qing Jiang[1],

Eran Greenberg[4], Stella Chariton[4], Vitali B. Prakapenka[4], Alexander F. Goncharov[1,2]

[1] Key Laboratory of Materials Physics, Institute of Solid State Physics, HFIPS, Chinese Academy of Sciences, Hefei 230031, China

[2] Earth and Planets Laboratory, Carnegie Institution of Washington, 5251 Broad Branch Road NW, Washington, DC 20015, USA

[3] Department of Mathematics, Howard University, Washington, DC 20059, USA

[4] Center for Advanced Radiations Sources, University of Chicago, Chicago, Illinois 60637, USA

Correspondence should be addressed to: agoncharov@carnegiescience.edu

**Details about the Methods used in this work**

Diamond anvil cells (DAC) of BX90 type [1] equipped with Boehler-Almax type seats and conical diamond anvils [2] and of symmetric type DAC with standard-design seats and diamond anvils (Almax- EasyLab) were used for single-crystal/powder diffraction and Raman measurements, respectively.

The double-sided laser-heating systems of the beamline GSECARS (APS, Sector 13, Argonne, USA) [3] features the flat top focal spot of 10 μm in diameter. The sample temperature was measured radiometrically (gray body approximation). The thermal radiation was recorded with a Princeton grating spectrometer (300 mm focal length) combined with PIXIS and PiMAx CCD array detectors.

The double-sided laser heating system at ISSP (Hefei) is combined with the Raman confocal system thus making concomitant Raman probing of the heating area very convenient [4]. Another similar Raman system coupled to a continuous flow He/$N_2$ cryostat was used for Raman experiment, where the sample was investigated upon unloading at room and low (78 K) temperatures [5].

The full description of the confocal GSECARS Raman systems, which has been used concomitantly to XRD measurements, has been published elsewhere [6]. It features five excitation laser lines from the UV to near IR, which can be automatically changed, and double-sided laser heating.

**Single-crystal X-ray diffraction analysis**

For the analysis of the single-crystal diffraction data (indexing, data integration, frame scaling and absorption correction) we used the *CrysAlis$^{Pro}$* software package. To calibrate an instrumental model in the *CrysAlis$^{Pro}$* software, *i.e.*, the sample-to-detector distance, detector's origin, offsets of goniometer angles, and rotation of both X-ray beam and the detector around the instrument axis, we used a single crystal of orthoenstatite (($Mg_{1.93}Fe_{0.06}$)($Si_{1.93}$, $Al_{0.06}$)$O_6$, *Pbca* space group, *a* = 8.8117(2), *b* = 5.18320(10), and *c* = 18.2391(3) Å). Using the Olex2 crystallography software package, the structures were solved with the ShelXT structure solution program [7] using Intrinsic Phasing and refined with the ShelXL [8] refinement package using Least Squares minimization. The powder diffraction images were integrated to powder patterns with Dioptas software [9]. Le-Bail fits of the diffraction patterns were performed with the Jana2006 software [10]. CSD 2022907 contain the supplementary crystallographic data for this paper ($K_3(N_2)_4$ structure). These data can be obtained free of charge from FIZ Karlsruhe *via* www.ccdc.cam.ac.uk/structures

**Structural analysis of cold-compressed $KN_3$**

Most of the cold-compressed $KN_3$ sample produces weak powder diffraction pattern as shown on the Figure S2. However, we were able to locate the grain that produces poor-quality single-crystal diffraction pattern. The diffraction patterns collected at the positions of the powder-like and single-crystal samples demonstrate that the same phase is present in both collection spots. The peak positions are only slightly different, which can be explained by pressure gradients in the sample without a pressure-transmitting medium. The differences in the distribution of intensities are clearly related to the strong preferred orientation of the single-crystal sample.

The unconstrained lattice determination from the single-crystal XRD dataset reveals the *I*-centered lattice with $a$ = 5.425(13), $b$ = 5.487(11), $c$ = 4.67(4) Å, $\alpha$ = 90.6(4)°, $\beta$ = 92.2(4)°, $\gamma$ = 90.22(17)°. These parameters are close to the parameters expected for the *I*4/*mcm* KN$_3$ compressed to ~50 GPa, however the structure is clearly distorted. The refinement of the *I*4/*mcm* KN$_3$ structure using our single-crystal dataset resulted in the good agreement factor R$_1$ = 11.03 % for 32/5 data to parameter ratio. This agreement demonstrates that we indeed deal with the distorted *I*4/*mcm* structure rather than with a completely different structure type, like predicted *C*2/*m* KN$_3$. The unconstrained lattice refinement suggests either monoclinic or triclinic symmetry. By a group-subgroup transformation we have generated the *I*2/*c* model of KN$_3$ from the high-symmetry *I*4/*mcm*. The refinement converged with slightly worse $R_1$ = 13.73 % with 56/8 data/parameter ratio. Although this value is higher than for the *I*4/*mcm* model, the use of monoclinic symmetry is justified by the lattice metrics. We provide the structure model of *I*2/*c* KN$_3$ below (Table S1) with a notice that the quality of the dataset allows to distinguish main structural motifs, while it should not be used for the judgement about specific structure details. For example, although the refined N-N distances in the azide group are reasonable (1.17(1) Å) they should be used with caution when considered in systematic comparison with other azides at similar conditions. Similarly, further symmetry reduction to triclinic does not provide additional reliable structural information.

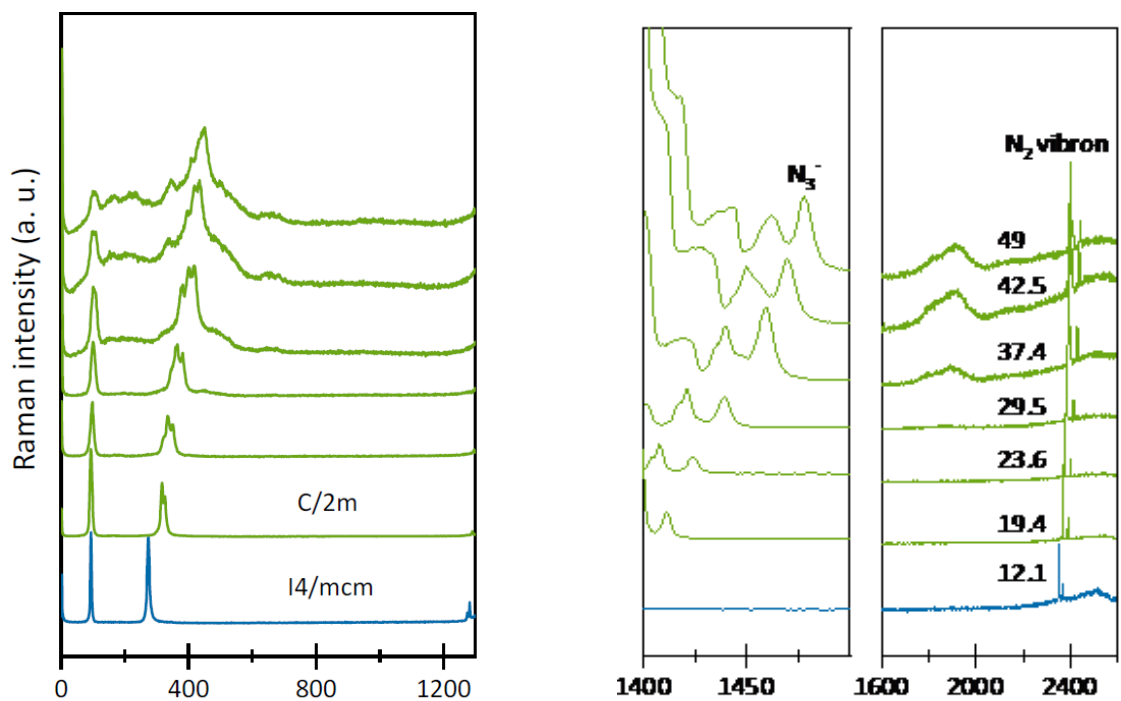

**Figure S1.** Raman spectra of KN$_3$ azide at various pressures (GPa) upon compression to 49 GPa. The excitation wavelength is 532 nm. The pressure medium is molecular nitrogen N$_2$, which yields the N-N stretch (vibron) spectra at about 2400 cm$^{-1}$.

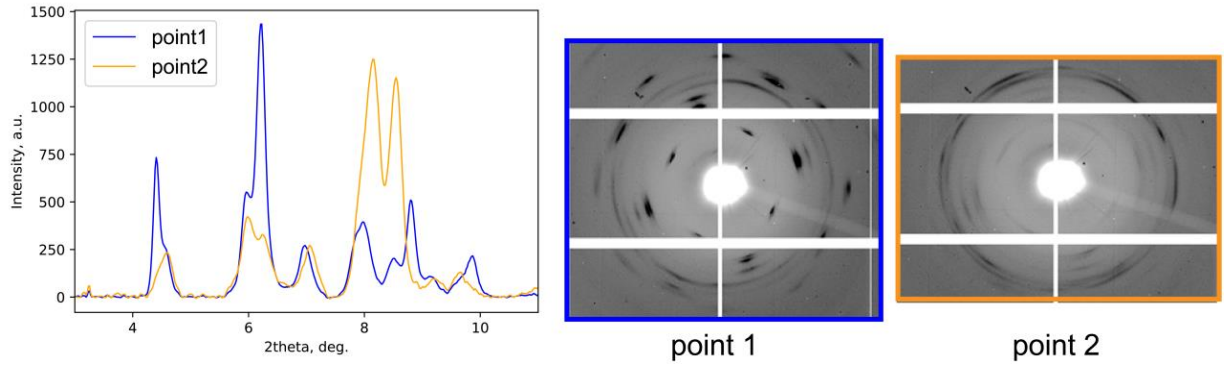

**Figure S2**. Integrated diffraction patterns and raw diffraction images of $KN_3$ at ~50 GPa collected at different parts of the sample. The X-ray wavelength is 0.2952 Å.

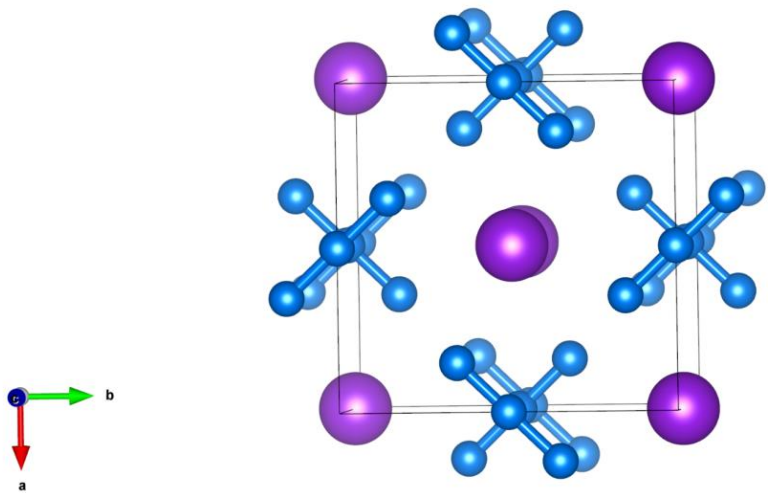

**Figure S3**. The crystal structure of $I2/c$ KN$_3$ at ~50 GPa.

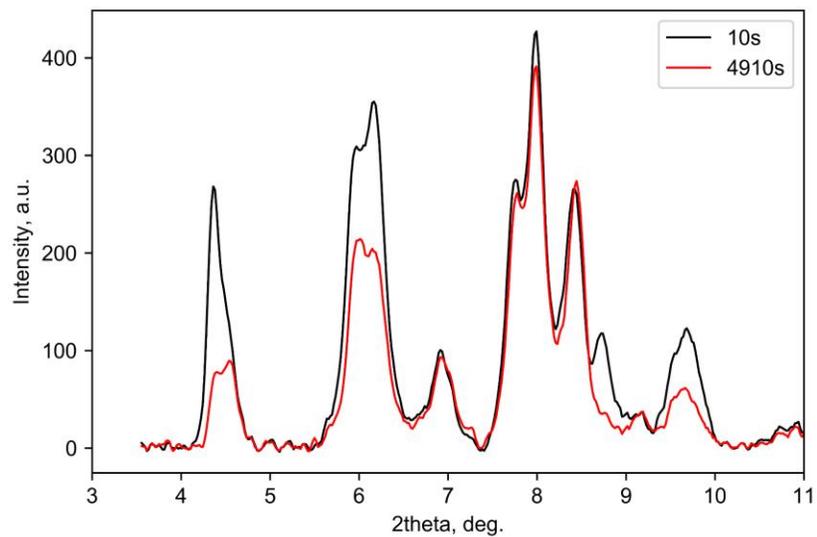

**Figure S4.** X-ray diffraction of KN$_3$ azide compressed to 50 GPa at room temperature collected after 10 and 4910 s of the X-ray exposure.

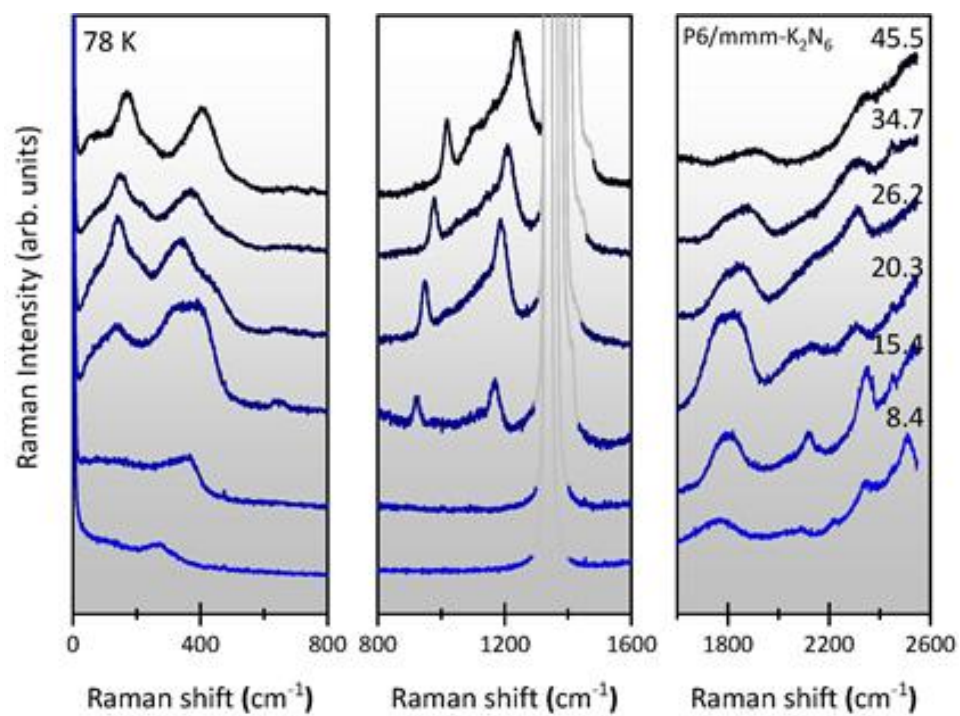

Figure S5. Raman spectra of $K_2N_6$ with hexazine rings at 80 K at various pressures upon pressure release.

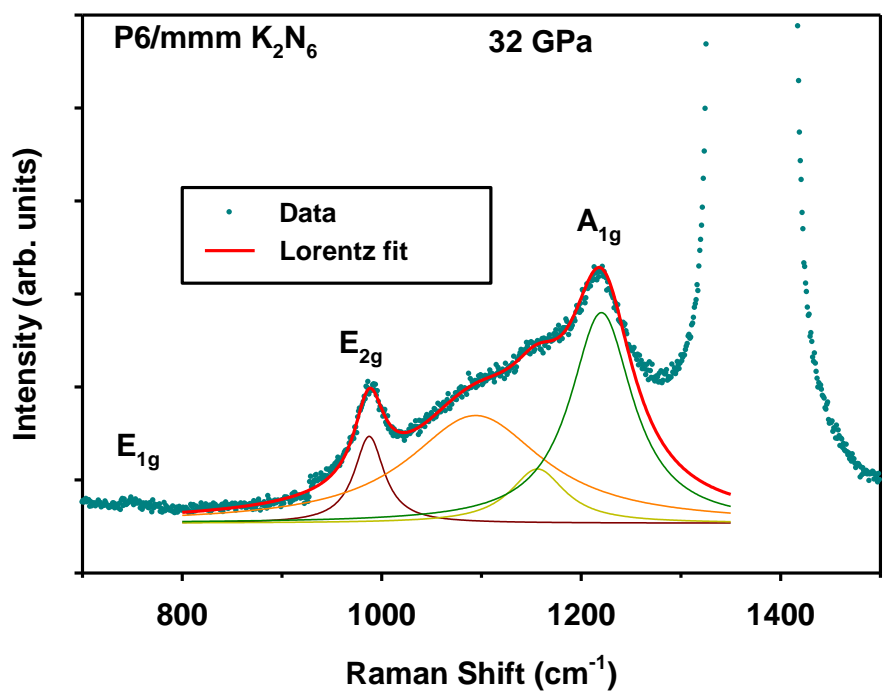

Figure S6. Raman spectra of $K_2N_6$ at 32 GPa and 297 K demonstrating the characteristic peaks of hexazine ring in the range of N-N stretching and stretching/deformation modes. The best fit to the data using four Lorentz profiles demonstrates the presence of additional bands assigned to disorder induced nominally IR active vibrations (see text) .

**Table S1. Crystal structure details of *I*2/*c* KN$_3$ at 50 GPa.**

| Space group | *I*2/*c* |
|---|---|
| Lattice parameters (constrained to monoclinic) | $a$ = 5.418(14), $b$ = 5.573(11), $c$ =4.89(4) Å, $\beta$ = 92.2(4)° |
| Atomic coordinates | K (0, 0.00059, 0.25) <br> N1 (0.150005, 0.65072, 0.00127) <br> N2 (0, 0.5, 0) |

**Table S2. Experimental and crystal structure details for $K_3(N_2)_4$ at 27 GPa**

| Crystal data | |
|---|---|
| Chemical formula | $K_3N_8$ |
| $M_r$ | 229.38 |
| Crystal system, space group | Tetragonal, $I4_1/amd$ |
| Temperature (K) | 300 |
| Pressure (GPa) | 27 (2) |
| $a, c$ (Å) | 5.331(2), 17.552(6) |
| $V$ (Å$^3$) | 498.8 (5) |
| $Z$ | 4 |
| Radiation type | Synchrotron, $\lambda$ = 0.29521 Å |
| $\mu$ (mm$^{-1}$) | 0.24 |
| Crystal size (mm) | $0.002 \times 0.002 \times 0.002$ |
| **Data collection** | |
| Diffractometer | GSECARS, 13 IDD, APS, USA |
| Absorption correction | Multi-scan *CrysAlis PRO* 1.171.40.75a (Rigaku Oxford Diffraction, 2020) Empirical absorption correction using spherical harmonics, implemented in SCALE3 ABSPACK scaling algorithm. |
| $T_{min}, T_{max}$ | 0.214, 1.000 |
| No. of measured, independent and observed [$I > 2\sigma(I)$] reflections | 365, 128, 87 |
| $R_{int}$ | 0.113 |
| $(\sin \theta/\lambda)_{max}$ (Å$^{-1}$) | 0.623 |
| **Refinement** | |
| $R[F^2 > 2\sigma(F^2)]$, $wR(F^2)$, $S$ | 0.074, 0.213, 1.16 |
| No. of reflections | 128 |
| No. of parameters | 12 |
| $\Delta\rho_{max}, \Delta\rho_{min}$ (e Å$^{-3}$) | 0.91, -1.36 |

| Crystal structure | | | | |
|---|---|---|---|---|
| Atoms | K1 | K2 | N1 | N2 |
| Wyckoff site | 8$e$ | 4$b$ | 16$f$ | 16$h$ |
| Atomic coordinates | | | | |
| $x$ | 0.5 | 0 | 0.8920(16) | 0.5 |
| $y$ | 0.75 | 0.75 | 0 | 0.3580(17) |
| $z$ | 0.46241(15) | 0.625 | 0.5 | 0.3808(4) |
| $U_{iso}$, Å$^2$ | 0.0162(13) | 0.0166(16) | 0.016(2) | 0.017(2) |